%% file: metamorph_fin.tex
\documentstyle[]{mn}
\input{epsf}
\input{mnfonts.tex}
\begin{document}

%%%%%%%%%%%%%%%%%%%%%%%%%%%%%%%%%%%%
\newcommand{\beq}{\begin{equation}}
\newcommand{\beqn}{\begin{eqnarray}}
\newcommand{\eeq}{\end{equation}}
\newcommand{\eeqn}{\end{eqnarray}}
\newcommand{\k}{{\kappa}}
\newcommand{\etal}{{\it et al.~}}

\title[A late-time transition in the cosmic dark energy?]
{A late-time transition in the cosmic dark energy?}

\author[Bruce\ A.\ Bassett, Martin\ Kunz, Joseph\ Silk and Carlo\ Ungarelli]
       {
        Bruce A.\ Bassett,$^1$\thanks{
        E-mail: bruce.bassett@port.ac.uk;
                        kunz@astro.ox.ac.uk; silk@astro.ox.ac.uk;
        carlo.ungarelli@port.ac.uk}
    Martin Kunz$^2$, Joseph Silk$^2$
        and Carlo Ungarelli $^1$ \\
        $^1$Institute of Cosmology and Gravitation,
            University of Portsmouth, 
            Mercantile House, Portsmouth, PO1 2EG, U.K. \\
        $^2$NAPL, Department of Physics, Keble Road,
            University of Oxford,
            Oxford, OX1 3RH, UK\\
       }

\date{
Accepted.
Received; in original form 2001 ...}

\pubyear{2001}

\label{firstpage}

\maketitle

\begin{abstract}
We study constraints from the latest CMB, large scale structure 
(2dF, Abell/ACO, PSCz) and SN1a data on dark energy models with a
sharp transition in their equation of state, $w(z)$. Such a
transition is motivated by models like vacuum metamorphosis
where non-perturbative quantum effects are important at late times. 
We allow the transition to occur at a specific redshift, $z_t$, to a final
negative pressure $-1 \leq w_f < -1/3$. 
We find that the CMB and supernovae data, in particular, prefer a
late-time transition due to the associated delay in cosmic
acceleration. The best fits ($\pm 1 \sigma$ errors) to all the data are
$z_t = 2.0{}^{+2.2}_{-0.76}$, $\Omega_Q = 0.73{}^{+0.02}_{-0.04}$ 
and $w_f = -1^{+0.2}$. 
\end{abstract}

%] \pacs{PACS numbers: }
 \noindent
%*
\section{Introduction}
The idea that the universe is currently accelerating comes from a
number of high-quality, but indirect, experiments. The luminosity distance
estimated from type Ia supernovae (Riess \etal 1998) favours 
recent acceleration
while the recent Cosmic Microwave Background (CMB) data (Netterfield \etal 2002, 
Lee \etal 2001, Halverson \etal 2002, Padin \etal 2001) 
suggest the universe has almost zero spatial curvature. 
This, combined with clustering estimates giving $\Omega_m \sim 0.3$ 
(Hamilton and Tegmark 2002), provides compelling
evidence for a dominant, unclustered, universal element; a conclusion
supported by the height of the first, and position of the second,
acoustic peak in the CMB (Kamionkowski \& Buchalter 2000).

Nevertheless, our understanding of the true nature of such an unclustered 
``ether" is arguably even worse than it is for the dark matter 
responsible for galaxy clustering. The oldest
idea is that it is a cosmological constant, $\Lambda \neq
0$.  This requires generating a tiny scale $\Lambda 
\sim (10^{-3} {\rm eV})^{4}$ and there have been many recent ideas to
achieve this (see e.g. Dienes 2001, Deffayet \etal 2001, 
Verlinde \& Verlinde 2000, Verlinde 2000), mostly motivated
by the revolution associated with higher dimension brane world models.

Undoubtedly the best-studied explanation, however, is quintessence 
(Ferreira \& Joyce 1998, Caldwell {\em et al} 1998); a very light scalar 
field $Q$, whose effective potential $V(Q)$ leads to acceleration in the 
late universe. However, quintessence
suffers from extreme fine-tuning since not only must one
set the cosmological constant to zero but one must arrange for
the quintessence field to dominate at late times only. This coincidence
problem  typically requires severe fine-tuning in the potential, although
this can be alleviated in models where the quintessence field couples to dark
matter (Holden \& Wands 2000, 
Amendola 2001, Tocchini-Valentini \& Amendola, 2001).

Such couplings to standard model fields imply extra fine-tunings
however since the quintessence field is extraordinarily light and
typically has Planck-scale vacuum expectation value today
$Q \sim M_{pl}$. If one believes that the quintessence field comes from
supergravity, then non-renormalisable
couplings to standard matter fields should appear automatically, 
{\em even if} all the {\em renormalisable} couplings to standard matter are 
fine-tuned to vanish.  

These non-renormalisable couplings, such as $Q
F_{\mu \nu}F^{\mu\nu}/M_{pl}$, where $F_{\mu\nu}$ is the Maxwell tensor,
cause variations of the fundamental constants of nature, such as the
fine-structure constant (Carroll 1998), and  the corresponding coefficients
must be made small to avoid conflicting with evidence at low redshift.
Similar constraints come from neutrino data (Horvat 2000) and make the
quintessence scenario appear extremely fine-tuned without some more
fundamental theoretical basis. 

Quite another possibility is that quantum effects have become
important at low redshifts and have stimulated the universe to
begin accelerating. Examples are {\em vacuum metamorphosis}, put forward
recently by Parker \& Raval (1999,2000: hereafter PR) and the work of 
Sahni and Habib (1998).

In particular, PR consider a scalar field of mass $m$ in a flat, FLRW
background,  and non-perturbatively compute the effective action 
in terms of the Ricci scalar, $R$. They
show that the trace of the  Einstein equations receives
quantum corrections some of which are proportional to
\begin{equation}
\frac{\hbar G m^4 R}{m^2 + (\xi - 1/6)R}[1 + {\cal O}(R)] .
\end{equation}
which  diverges when $R \rightarrow -m^2/(\xi - 1/6)$, 
signalling  significant quantum contributions to the equation of
state of the scalar field. Here $\xi \neq 1/6$ is the non-minimal 
coupling constant. At early times the equation of
state is dust on average \footnote{A massive scalar field acts
like dust on averaging over many oscillations of the field. Since
the scalar field here is so light this may be a bad approximation
since the period of oscillation is so long.} and then makes a
transition from dust to cosmological constant plus radiation (PR
1999).

To explain the supernova Type Ia (SN1a) data the scalar 
field is forced to be extremely light, 
$m^2/(\xi - 1/6) \sim 10^{-33}$ eV. Vacuum
metamorphosis therefore suffers from the same fine-tuning problems
as quintessence - why are there no dimension 5 operators leading
to unacceptable variation of fundamental constants?

The idea of a sudden phase transition is very attractive however
and is more general than just the example of vacuum metamorphosis. 
In fact the idea of late-time phase transitions is rather old, dating 
back at least as far as 1989 (Hill {\em et al} 1989, 
Press {\em et al} 1990).

We therefore choose a phenomenological model which captures the
basic features of a phase transition in the equation of state, but
which is not strictly linked to any specific model. We then ask
whether current CMB and large scale structure (LSS) data rule out
such a transition, or indeed, favour it over the now standard
$\Lambda$CDM model.
%%%%%%%%%%%%%%%%%%%%%%%%%%%%%%%%%%%%%%%%%%%%
\section{The Phenomenological Model}
%%%%%%%%%%%%%%%%%%%%%%%%%%%%%%%%%%%%%%%%%%%
In addition to baryons, neutrinos and cold dark matter our
model is characterized by a scalar field $Q$ with a redshift
dependent equation of state  $p_Q = w(z) \rho_Q$. We 
choose $w(z)$ to have the following
form 
%%%%
\beq w(z) = w_0 +  \frac{(w_f - w_0)}{1 + \exp(\frac{z -
z_t}{\Delta})} \label{w} \eeq
%%%%
In this paper, we shall restrict ourselves to the case where the
initial equation of state is pressure-free
matter, $w_0 = 0$. In terms of the scalar field dynamics, we will consider a
class of models characterized by three free parameters: (1) the
final value of the equation of state, $w_f$, (2) the redshift
$z_t$ of the transition, and (3) the energy density $\Omega_Q$ of the scalar
field in units of the critical energy density $\rho_{crit}$; 
$\Omega_Q=\rho_Q/\rho_{crit}$ \footnote{Notice that the width of the transition
is controlled by $\Delta$. Double precision limits
require that $z_T/\Delta < 100$ while the constraint that $w =
w_f$ at $z = 0$ implies that $z_T/\Delta > 10$. To satisfy both
of these constraints simultaneously for $0 < z_T < 2000$, we have
chosen a $z_t$-dependent $\Delta$ defined by  $z_T/\Delta \equiv
30$.}

We also assume that any coupling of the scalar field with other
fields are negligible. In this case the energy density
$\rho_Q$ is determined from energy conservation
\begin{equation}
\dot{\rho}_Q = -3H\rho_Q(1 + w(z))
\label{energyconserve}
\end{equation}
which can be explicitly integrated since $w(z)$ and $\Omega_Q$ today are 
given. Using Eq.~(\ref{w}) for $w(z)$ and specifying initial conditions for the
scalar field one obtains the scalar field potential $V(Q)$ and its
derivatives along the ``background'' trajectory $Q(t)$. In
particular, $V^{\prime}(Q)$, is
\begin{equation}
V^{\prime}(Q)=-\frac{3H}{2}\sqrt{(1+w)\rho_Q}\left[1-\frac{1}{3}
(1+z)\frac{d\log(w(z))}{dz}\right]
\label{vprime}
\end{equation}
and $V^{\prime\prime}(Q)$ can be easily obtained  from
Eq.~(\ref{vprime}). $V^{\prime\prime}(Q)$ is required to solve the
evolution equation for the scalar field fluctuations, $\delta
Q({\bf x},t)$. In the synchronous gauge, the Fourier modes $\delta
Q_k(t)$ obey the equation
\beq \delta \ddot{Q}_k + 3H \delta\dot{Q}_k +
\left(\frac{k^2}{a^2} + V^{\prime\prime}\right)\delta Q_k = - \dot{h}_k \dot{Q}
\label{ddotq} \eeq
where $h$ is the trace of the spatial metric perturbation~ (Ma and
Bertschinger 1995). We choose adiabatic initial conditions $\delta
Q_k=\delta\dot{Q}_k=0$ for the scalar field.

\section{The data and analysis pipeline}

\subsection{The cosmic parameters}

Due to computational restrictions we fixed cosmic parameters
not directly linked with the scalar field $Q$. We therefore
performed a likelihood analysis in the neighbourhood of the best
fit standard model. Hence 
we have no assurance that a better global minimum for
$\chi^2$ does not exist. However, our main goal is to test whether
the current data favour a phase transition over the standard
$\Lambda$CDM model, and for this our analysis is sufficient. We
chose the following ``standard'' cosmic parameters (Wang \etal 2001):

\begin{itemize}
\item $H_0 = 65$ km/s/Mpc
\item $\Omega_b = 0.05$, $\Omega_{\rm tot} = 1$ (flat universe)
\item $n_s = 1$, $n_t = 0$,  $\tau = 0$\,.
\end{itemize}
Here $\tau$ is the reionisation optical depth and $n_{s,t}$ are
the spectral indices for scalar and tensor perturbations
respectively. We set $\Omega_{\Lambda} = 0$ and included only the 
effective $3.04$ massless standard model neutrinos. We did not include
tensor perturbations and 
as we varied $\Omega_Q$ today we specified  the cold dark matter density
to ensure overall flatness of the universe, {\em viz.}
$\Omega_{\rm CDM} \equiv 1 - \Omega_{b} - \Omega_Q$. We 
emphasize here that fixing $\Omega_{b}$ and $H_0$ can produce
artificially narrow likelihood curves especially for $\Omega_Q$
and that our results should not be interpreted as determining the
true energy density of the scalar field to such a precision.

\subsection{Observational Data and Analysis}

We constrain the parameters of our phenomenological model by
comparing its predictions with a number of observations.

\underline{\bf CMB}: we use the decorrelated COBE DMR data (Tegmark \& Hamilton 
1997) to constrain the fluctuations on large scales and
combine it with the recent data from the BOOMERanG (Netterfield
\etal 2002), MAXIMA (Lee \etal, 2001) and DASI (Halverson \etal 2002)
experiments. In total we used $49$ data points ranging in $\ell \in
[2,1235]$. We take into account the published
calibration uncertainties for BOOMERanG, MAXIMA and DASI but not
the pointing and beam uncertainties, since they are not public
for all experiments.

\underline{\bf LSS}: we use the matter power spectrum inferred from the 2dF
100k redshift survey (Tegmark, Hamilton \& Xu 2001), the
IRAS PSCz 0.6 Jy survey (Hamilton \& Tegmark, 2000) and the Abell/ACO cluster
survey (Miller \etal 2001). We limit the comparison to
$k < 0.2 h/{\rm Mpc}$ to minimise potential non-linear contaminations.
All together, we use $48$ points. We do not currently 
include the Ly-$\alpha$
analysis of Croft et al (2000). Even though our best fitting
models agree quite well with the shape of the recovered linear
matter power spectrum, the concerns raised by (Zaldarriaga \etal 2001) 
make it seem preferable to postpone the use of this data
set.

\underline{\bf SN1a}: we use the redshift-binned 
supernova data from Riess \etal (2001), which
includes the HZT (Riess \etal 1998) and SCP (Perlmutter \etal 1999) data.

We follow the standard approach of computing the $C_{\ell}$'s and $P(k)$ for
each set of parameters over the 3 dimensional grid $(\Omega_Q,
z_T, w_f)$, using a modified version of
CMBFAST (Seljak \& Zaldarriaga 1996). We then evaluate the corresponding
$\chi^2$ at each grid point.

We find that the likelihood values for the CMB depend slightly
on the likelihood functional used, eg. a simple 
$\chi^2$ computed using the $C_{\ell}$'s or an offset 
log-normal distribution (Bond \etal 2000), but we are not certain
if this is an intrinsic difference, or due to the
slightly different data in the RADPACK package. However the overall changes
are within $1 \sigma$ and hence not significant.

The link between CMB and LSS is given by the respective
normalisations. The CMB data fixes the overall amplitude of the
model quite precisely. As the connection between the matter power
spectrum inferred from galaxy and cluster surveys and the actual
distribution of dark matter is much less clear, we allowed a bias,
$b \in (1/5, 5)$ for 2dF and PSCz and $b \in (1/9,9)$ for
Abell/ACO since clusters are expected to be more biased than
galaxies. Here $b$ is the factor between the perturbation
amplitudes, and hence enters quadratically in the power spectra.

In general we marginalise over parameters by integrating the likelihood.
We find that the results are consistent with those found from maximising the 
likelihood. This is expected for a nearly Gaussian likelihood, and 
it provides some reassurance that the $\chi^2$ method is justified.

\section{The physics of metamorphosis}

\subsection{The CMB}

To understand the imprint of metamorphosis on the $C_{\ell}$'s of
the CMB requires two insights. First the contribution of
the scalar field to the expansion rate of the universe is
negligible for $z > 3$ if $w_f < -0.4$ since
$\rho_Q \sim a^{-3(1 + w_f)}$. This implies that the 
dynamics of $Q$ has little effect on the
evolution of the metric perturbations (which respond to the total
matter perturbation) and hence the $C_{\ell}$'s are almost
insensitive to transitions with $z_t > 3$.

%%%%%%%%%%%%%%%%%%%%%%%%%%%%%%%%%%
\begin{figure}
\epsfxsize=3.4in 
\epsffile{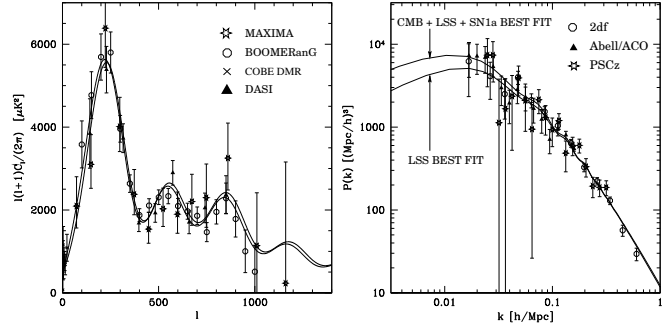}
\caption{{\bf Left}: The total-data best-fitting model (right curve) vs 
$\Lambda$CDM (left curve) which both have $\Omega_Q = 0.73$ and 
$w_f = -1$ but $z_t = 1.5$ for the best-fit. 
{\bf Right}:The power spectrum for our total best-fit model with 
$(\Omega_Q, z_t,
w_f) = (0.73, 1.5,-1)$ compared with the LSS best-fit ($0.7, 6.5,-0.55)$. The 
LSS data shown are the linear transfer functions inferred from the 2df and 
PSCz galaxy surveys and the Abell/ACO cluster survey. We do not show the 
lyman-$\alpha$ data. } \label{clpk}
\end{figure}
%%%%%%%%%%%%%%%%%%%%%%%%%%%%%%%%%%

\begin{figure}
\epsfxsize=3.4in
\epsffile{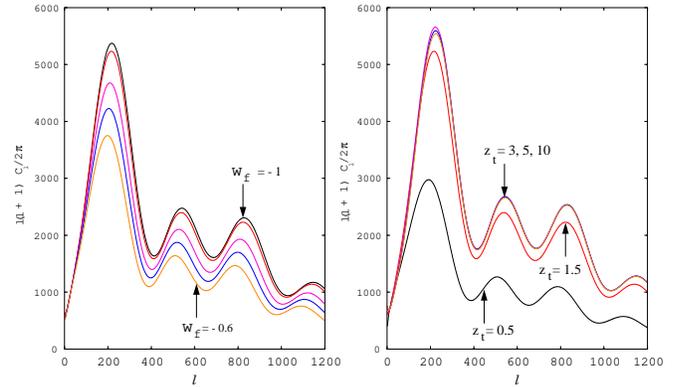} \caption{Variation of the $C_{\ell}$'s with $z_t$ 
and $w_f$. {\bf Left}: The $C_{\ell}$ curves increase monotonically with 
decreasing $w_f$ starting with  $-0.6$ (bottom), 
$-0.7, -0.8,-0.95$ and ending with $-1$ (top). The primary change is in the 
normalisation of the spectrum through the change in the ISW effect. 
{\bf Right:} The $C_{\ell}$'s for $z_t = 0.5$ (bottom), $1.5, 3,5$ and $10$.
The ISW contribution to the COBE normalisation changes very rapidly for 
small $z_t$ which allows for {\em delayed acceleration}. However, 
the CMB is insensitive to $z_t > 3$ since the scalar 
field is dynamically irrelevant at those redshifts. } 
\label{cl_ztwf}
\end{figure}

This is evident in Fig. (\ref{cl_ztwf}). The figure also shows
an effect which at first sight is perhaps surprising:  the CMB is
extremely sensitive to $z_t$ for $z_t < 3$. This is clarified once
we remember that the standard $\Lambda$CDM CMB has a large
Integrated Sachs-Wolfe (ISW) contribution due to the decay of the
gravitational potential $\Phi$ during the epoch of acceleration,
typically occurring at $z < 2$.

Therefore, if we choose two models with $w_f < -0.5$ but with 
$z_t = 0.5$ and $z_t = 1.5$ respectively, the initiation of 
the accelerated phase will vary by around
$300\%$, and the decay of the gravitational potential starts
at very different epochs. Hence the fluctuations in the CMB 
for these two models differ mainly on large angular scales which
alters the acoustic-peak/SW-plateau ratio.

In addition, $z_t < 2$ implies that the rapid change in $w$ can
cause its own effect on the gravitational potential since the
energy density of $Q$ is starting to dominate. This gives rise to
an ISW effect purely due to the  scalar field {\em dynamics}.
We will see later that these two effects almost completely explain
the behaviour of the likelihood curves in  Fig.(\ref{1dlike}) which
are very sensitive to small $z_t$ but exhibit a long tail for
large $z_t$.

The other parameter of interest is $w_f$. This has a simple effect
on the CMB. As $w_f$ decreases towards to $-1$ the universe starts
to accelerate earlier and is accelerating more violently today.
This alters the ISW effect which is important on large scales and
which contributes to the COBE normalisation. The effect of $w_f$
is mainly then to amplify or supress the $\ell > 50$ $C_{\ell}$'s
by a more-or-less $\ell$-independent amount. This is evident in
Fig. (\ref{cl_ztwf}).

\subsection{The matter power spectra}

We now discuss the effect of $z_t$ and $w_f$ on the CDM and $Q$
power spectra. A key point is that the scalar field is very light both
before and after the transition at $z_t$. This means that the
associated Compton wavelength, $\sim (V^{\prime\prime})^{-1/2}$, of the $Q$
field is very large (as in standard quintessence models) for all
times. This implies that no clustering occurs in the $Q$ field on small
scales ($k > 0.01 h$ Mpc$^{-1}$).

After the transition the potential $V(Q)$ becomes even flatter (in
order to obtain acceleration) and hence the Compton wavelength
increases. This means that clustering now only occurs on the
largest scales for $z < z_t$. This allows us an 
intuitive idea of the effect of $z_t$. As $z_t$
is increased, we see that the Compton wavelength effect forces
clustering to occur only on larger and larger scales. However,
this is actually only a fairly weak effect and the dominant
variable is $w_f$.

The effect of $w_f$ on the $Q$ power spectrum is straight-forward:
for fixed $z_t$, the closer $w_f$ is to $-1$, the less clustering
occurs. Conversely, the closer $w_f$ is to $0$, the more
clustering occurs. This is clear in the quintessence case from the
work of Ma \etal (1999) which can be understood by 
rewriting the RHS of equation
(\ref{ddotq}) in terms of the CDM density perturbation $\delta_c
\equiv \delta \rho_{CDM}/\rho_{CDM}$ which becomes 
$\delta_c [(1 + w_f)\rho_Q]^{1/2}$. Clearly this driving
term drops to zero as $w_f \rightarrow -1$.

For fixed $w_f$, increasing $z_t$ implies that the universe
spends {\em less} time in the dust phase where $w = 0$ and hence
the long wavelengths of $\delta Q$ have less time to grow relative
to the short wavelengths (that will not grow irrespective of the
values of $z_t$ and $w_f$ ). This effect is clearly visible in
Fig.~\ref{clpk} where
we show the global best-fit CDM power spectrum with $z_t = 1.5$ 
together with the best-fit to just the LSS data which has $z_t = 6.5$ and 
hence less power on large scales. Similarly for fixed $z_t$, increasing 
$w_f$ towards zero allows more clustering on large scales relative 
to small scales.

An important point is that simply specifying $w_f = -1$ and $z_t >
1000$ does not imply that the resulting model is the same as a
$\Lambda$CDM model. While the transfer function $T(k) = 1$ on all
scales for $z < z_t$ if $w_f = -1$, this does not mean the initial 
power spectrum $P_Q(k)$ was zero, 
whereas $P_{\Lambda}(k) \equiv 0$ since $\delta \Lambda = 0$ 
by definition. This is visible in the $C_{\ell}'s$ of the 
left panel of figure (\ref{clpk}). Both curves have $\Omega_Q = 0.73$ and 
$w_f = -1$. They differ due to the fluctuations $\delta Q$ and the 
transition at $z_t = 1.5$. 

This means that despite the background dynamics being essentially
equivalent for the two models for $z_t \rightarrow \infty$, 
the perturbations are not the same,
and indeed one could consider adiabatic or isocurvature initial
conditions for the $\delta Q$. Hence, simply showing that the recent
dynamics of the universe favours $w = -1$ does not of itself, prove
that the acceleration comes from the cosmological constant.

Tests sensitive to perturbations are also required. This may be
particularly important in the case when the scalar field is
non-minimally coupled to the spacetime curvature (see e.g.
Perrotta \& Bacciagalupi, 2001). 

\subsection{The SN1a data}\label{sndiscuss}

\begin{figure}
\epsfxsize=3.0in \epsffile{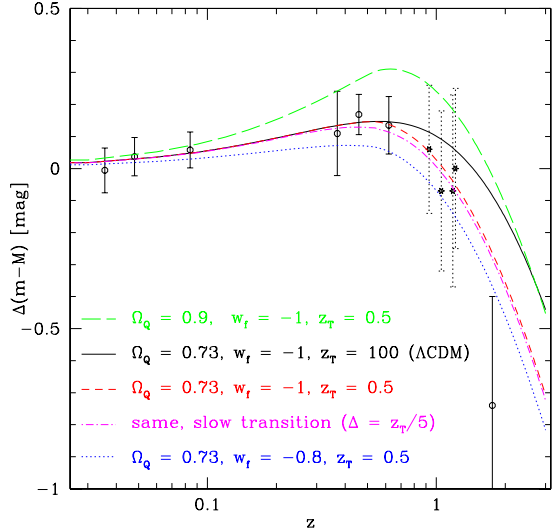} \caption{The
redshift dependence of the luminosity distance (as magnitudes)
minus an empty ($\Omega = 0$) universe for four different
metamorphosis models. The $z_t = 0.5$ model fits the data best, mainly 
due to the single data point at $z = 1.7$. 
The solid line effectively coincides with a $\Lambda$CDM
model. The redshift-binned SN1a data is from Riess et al (2001);
the four dashed data points are experimental and were not
included in the fit.} \label{sn1a}
\end{figure}

To compare with the supernovae type Ia measurements we compute the
luminosity distance:
\beq 
d_L(z) = c (1+z) \int_0^z du/H(u) 
\eeq
from which $\Delta(m-M)= 5[log_{10}(d_L(z))-log_{10}(d_{L_0}(z))]$
where $d_{L_0} = c z (z+2)/(2 H_0)$ is the empty-beam distance. 
We find that a step-function
approximation to $w(z)$ is extremely accurate, due to the integral
nature of the luminosity distance (see the dash-dotted line versus
the short dashed line in figure (\ref{sn1a})).

Figure (\ref{sn1a}) shows $\Delta(m-M)$ for a variety of models while
figures (\ref{1dlike}) and (\ref{2dlike}) show the results of the
likelihood analysis. Models with a very recent transition
from deceleration to acceleration are favoured since they fit the
highest $z$ supernova best, while still being consistent with the
intermediate-$z$ supernovae at $z \sim 0.5$. Due to the large
error bars, the constraints are weak, however. Furthermore, since the
metamorphosis models are almost indistinguishable from standard
$\Lambda$CDM models for $z < z_t$, we have only constraints on
$z_t$ less than the redshift of the farthest supernova observed
($z \approx 1.75$).

%%%%%%%%%%%%%%%%%%%%%%%%% SN1a FIG 
%%%%%%%%%%%%%%%%%%%%%%%%%%%%%%%%%%%%%%%%%%

The dependence on the other parameters is very much the same as
for conventional dark energy models. Not surprisingly, we recover
for $\Omega_Q$ vs $w_f$ the results of Turner and Riess (2001). 

\subsection{BBN constraints}

Big-bang nucleosynthesis (BBN) primordial abundances depend
sensitively on the expansion rate of the universe at the
temperature $T = 1$MeV which controls the neutron-proton ratio.
Assuming that the quintessence field scales as radiation at
nucleosynthesis Bean \etal (2001) set the 
limit of $\Omega_Q < 0.045$ at $2\sigma$ at $T = 1MeV$.

Since we assume the initial $w_{0}=0$, the scalar field scales as
dust for $z > z_t$ and hence $\Omega_Q$ is dynamically negligible
at nucleosynthesis which therefore provides no constraints on our parameters.

If we broadened our parameter set to include $w_{0}$ we would
expect BBN to set joint limits on the parameters. In particular,
for $w_{0} = 1/3$ (radiation), the BBN data would favour large
$z_t$, $w_f$ close to $-1$ and smaller $\Omega_Q$.

\section{Results}

\begin{figure}
\epsfxsize=3.4in
\epsffile{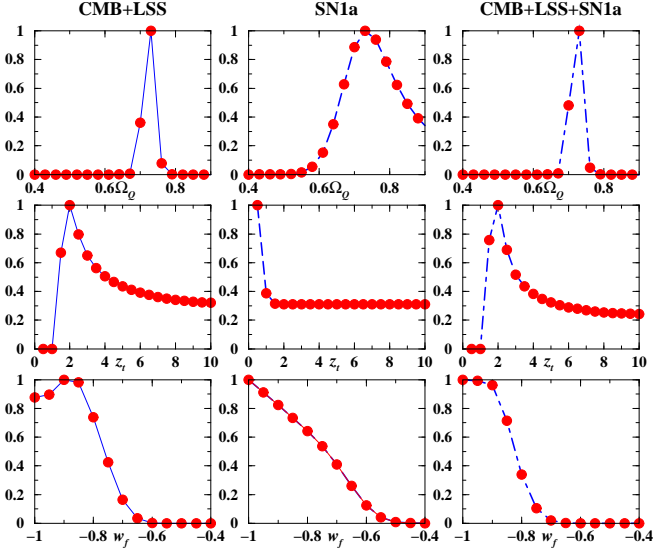} \caption{The marginalised 1-d likelihood
plots for our variables $(\Omega_Q, z_t, w_f)$. {\bf Left column}: CMB + LSS, 
{\bf middle column}: SN1a, {\bf right column}: Total data set. Both the CMB and
SN1a data favour a late transition (small $z_t$) due to the
corresponding delay in cosmic acceleration. We computed these
marginalised likelihoods using both integration and maximisation
and the results were similar, as they should be for Gaussian
likelihoods.}
\label{1dlike}
\end{figure}

\begin{figure}
\epsfxsize=3.4in
\epsffile{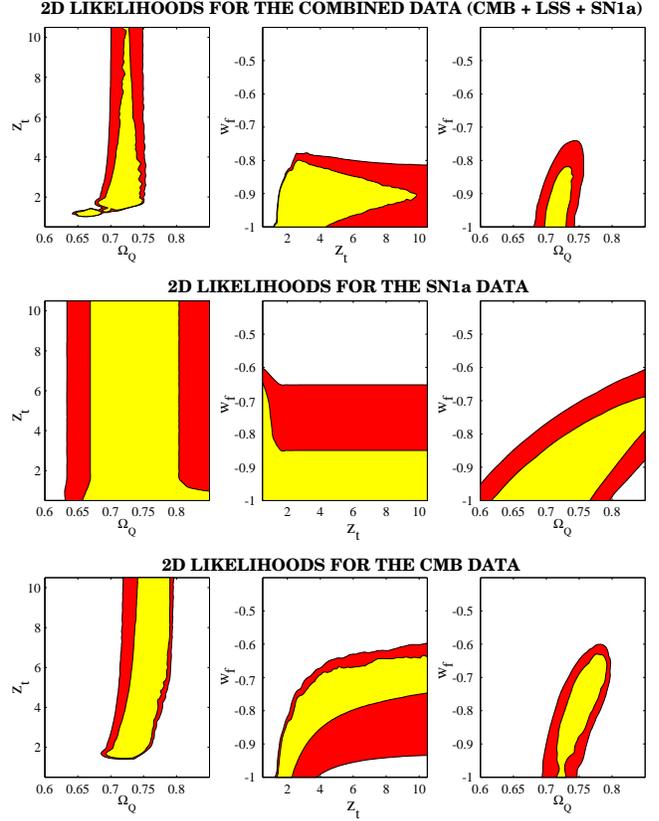} \caption{The marginalised 2-d
likelihood plots for the combined CMB, LSS and SN1a data sets
showing 1 and 2$\sigma$ contours defined as where  
the integral of the normalised
two-dimensional likelihoods are equal to $0.68$ and $0.95$
respectively.} \label{2dlike}
\end{figure}

Figures (\ref{1dlike}) and (\ref{2dlike}) show our main results
through the marginalised 1-d and 2-d likelihoods for
$(z_t,w_f,\Omega_Q)$.

We do not show the likelihoods for CMB and LSS alone since the LSS 
provides only very weak constraints on $z_t$ (slightly preferring
higher values) and $w_f$ (no constraints at all). It prefers an
$\Omega_Q$ around $0.7$, consistent with the CMB likelihood.
Due to these very weak results (and the consistent result for
$\Omega_Q$), the likelihood for the CMB data looks just like the
one for the combined CMB+LSS data.

The supernovae prefer (as explained in section \ref{sndiscuss})
a low $z_t$, but its significance (stemming from only one 
supernova at $z > 1$) is too weak to change the overall likelihood by much.
The constraints on $\Omega_Q$ are weaker than for the other data
sets, but consistent. Furthermore, the SN1a data prefers $w \approx -1$,
which tightens the overall constraints on the equation of state
somewhat, leading to $w < -0.8$ at the 1$\sigma$ level. 

In Figs. (\ref{clpk}) and (\ref{sn1a}) we show 
our best fits versus the current CMB, LSS and SN1a data and theoretical
predictions of the standard $\Lambda$CDM model.

The $\chi^2$ values of the overall best fit model (with $w_f=-1.0$,
$z_t = -1.5$ and $\Omega_Q = 0.73$) are $33$ (CMB) + $36$ (LSS)
+ $4$ (SN1a), in total $73$. On the same parameter grid, the best
fit $\Lambda$CDM model has $\Omega_Q = 0.73$ as well. Its $\chi^2$
values are $40$ (CMB), $34$ (LSS) and $4$ (SN1a), in total $78$.

We used $49$ data points for the CMB, $48$ for the LSS and $7$
for the supernova data. We allowed a free overall model normalisation
plus a bias/calibration uncertainty for each of the three LSS and
the four CMB data sets. Our phenomenological model has three
free parameters, while the $\Lambda$CDM models have only one.
So in total we have approximately (neglecting correlations within
the experiments as well as between them) $49+48+7-1-4-3-3=93$
degrees of freedom for our model, and $95$ dof for the $\Lambda$CDM 
models.

We can see that both groups of models are perfectly consistent
with current data. Given the error bars of the data sets, the family
of $\Lambda$CDM models is included in our phenomenological models
for $w = -1$ and large $z_t$. The figures show that current data
slightly prefers a low-$z$ phase transition, which is still true
when taking into account that we have to add two degrees of
freedom for pure $\Lambda$CDM models. On the other hand, the difference 
is too small to speak of a detection; assuming Gaussian
errors and 3 ``parameters of interest'' ($\Omega_Q$, $w_f$, $z_t$),
models with a $\Delta \chi^2$ of $5$ above the best-fit would 
formally be excluded at about 83\%, hence somewhere between 1 and 2 $\sigma$.

\section{Conclusions and tests for late transitions}

We have studied a phenomenological model in which the dark 
energy of the universe is described by a
scalar field $Q$ whose equation of state $w$ undergoes a sudden
transition ({\em metamorphosis}) 
from $w_0 = 0$ (dust) to $w_f < -0.3$ at a specific redshift $z_t$.

While similar to the quintessence paradigm in practical respects,
the underlying philosophy is very different since we are interested in 
the possibility of detecting radical physics in the dark energy, 
such as the vacuum  metamorphosis model (PR). We used the current CMB,
large-scale structure (LSS) and supernovae (SN1a) data to 
constrain our phenomenological parameter
space variables $(\Omega_Q, z_t, w_f)$.
 
The CMB and SN1a data are sensitive to a transition if it occurs 
at low redshifts ($z_t < 3$)  due to the delay in the epoch at which cosmic 
acceleration can begin, relative to the standard $\Lambda$CDM models. 
We found that
\begin{itemize}
\item The global best-fit to the current data occurs for $z_t = 1.5, 
w_f =-1.0$ and $\Omega_Q = 0.73$ while the marginalised 1d likelihood for 
$z_t$ peaks at $z_t = 2.0$. The best-fit model is consistent with
the data and is a marginally better fit than the best $\Lambda$CDM
model.
\item The CMB provides the best constraints on the parameters,
especially on $z_t$
due to the integrated Sachs-Wolfe effect (Fig. \ref{1dlike}).
\end{itemize}
Is it possible to distinguish this metamorphosis from standard 
quintessence models? 

This may be difficult since studies of quintessence favour similar values 
for $w_f$ and $\Omega_Q$ (Wang {\em et al} 1999, Bacciagalupi \etal 2002, 
Corasaniti \& Copeland 2002) and $c_s^2 = 1$ in both cases (Erickson \etal 2001). 
Traditional methods to discover variation in
$w$ based on the luminosity/area distance may be sufficient to discover 
transitions at $z_t < 2$ but may be statistically inefficient in
separating metamorphosis from standard quintessence since the distances
effectively depend on the integral $\int w(z) dz$ (Maor \etal 2001).

An interesting alternative (Jimenez \& Loeb 2001) is 
measurements of age differences in passively-evolving galaxies at
different redshifts which in principle allow direct determination of $w(z)$
and a direct test of metamorphosis if $z_t$ lies in the 
epoch of galaxy formation ($z_t < 5$).

Furthermore, future very high-precision measurements of the 
CMB at 1\% or better might be able to detect the perturbations in the microwave background
from the fluctuations in the scalar field itself, which would not
be present in a smooth background component like a cosmological constant 
though separating this out from lensing and foreground contamination will be
very difficult. 

Finally, an intriguing possibility is that the rapid transitions studied here
may provide a solution to the current impasse for quintessence
models in explaining the varying-$\alpha$ data,{\em viz}: quintessence
models can explain the apparent variation of $\alpha$ around $z
\sim 1-3$ but cannot then simultaneously match the results of
the Okun natural reactor (Chiba \& Khori 2001) at $z \sim 0$.
Detailed analysis of these issues is left to future work.

\section*{Acknowledgements}

We thank Luca Amendola, Carlo
Baccigalupi, Rachel Bean, Rob Caldwell, Takeshi Chiba, 
Pier Stefano Corasaniti,  Marian Douspis, Andrew
Hamilton, Steen Hansen, Rob Lopez, Roy Maartens, Max Tegmark and 
David Wands for useful
discussions on a variety of issues. We thank Chris Miller for
providing us with the Abell/ACO data, and Adam Riess for the
combined SNIa data.

BB acknowledges Royal Society support and useful discussions with 
the groups  at RESCEU, Kyoto, Osaka and Waseda.  
MK acknowledges support from
the Swiss National Science Foundation. CU is supported by the
PPARC grant PPA/G/S/2000/00115.

\end{document}

%% file: mnfonts.tex
\newif\ifAMStwofonts
\ifoldfss
  \ifCUPmtlplainloaded \else
    \NewTextAlphabet{textbfit} {cmbxti10} {}
    \NewTextAlphabet{textbfss} {cmssbx10} {}
    \NewMathAlphabet{mathbfit} {cmbxti10} {} % for math mode
    \NewMathAlphabet{mathbfss} {cmssbx10} {} %  "   "    "
  \fi
  \ifAMStwofonts
    \ifCUPmtlplainloaded \else
      \NewSymbolFont{upmath} {eurm10}
      \NewSymbolFont{AMSa} {msam10}
      \NewMathSymbol{\upi}     {0}{upmath}{19}
      \NewMathSymbol{\umu}     {0}{upmath}{16}
      \NewMathSymbol{\upartial}{0}{upmath}{40}
      \NewMathSymbol{\leqslant}{3}{AMSa}{36}
      \NewMathSymbol{\geqslant}{3}{AMSa}{3E}

      \let\leq=\leqslant 
       
    \fi
  \fi
\fi % End of OFSS

\ifnfssone
  \newmathalphabet{\mathit}
  \addtoversion{normal}{\mathit}{cmr}{m}{it}
  \addtoversion{bold}{\mathit}{cmr}{bx}{it}
  \newmathalphabet{\mathbfit} % math mode version of \textbfit{..}
  \addtoversion{normal}{\mathbfit}{cmr}{bx}{it}
  \addtoversion{bold}{\mathbfit}{cmr}{bx}{it}
  \newmathalphabet{\mathbfss} % math mode version of \textbfss{..}
  \addtoversion{normal}{\mathbfss}{cmss}{bx}{n}
  \addtoversion{bold}{\mathbfss}{cmss}{bx}{n}
  \ifAMStwofonts
    \ifCUPmtlplainloaded \else
      \UseAMStwoboldmath
      \makeatletter
      \new@mathgroup\upmath@group
      \define@mathgroup\mv@normal\upmath@group{eur}{m}{n}
      \define@mathgroup\mv@bold\upmath@group{eur}{b}{n}
      \edef\UPM{\hexnumber\upmath@group}
      \new@mathgroup\amsa@group
      \define@mathgroup\mv@normal\amsa@group{msa}{m}{n}
      \define@mathgroup\mv@bold\amsa@group{msa}{m}{n}
      \edef\AMSa{\hexnumber\amsa@group}
      \makeatother
      \mathchardef\upi="0\UPM19
      \mathchardef\umu="0\UPM16
      \mathchardef\upartial="0\UPM40
      \mathchardef\leqslant="3\AMSa36
      \mathchardef\geqslant="3\AMSa3E

      \let\leq=\leqslant 

    \fi
  \fi
\fi % End of NFSS release 1

\ifnfsstwo
  \DeclareMathAlphabet{\mathbfit}{OT1}{cmr}{bx}{it}
  \SetMathAlphabet\mathbfit{bold}{OT1}{cmr}{bx}{it}
  \DeclareMathAlphabet{\mathbfss}{OT1}{cmss}{bx}{n}
  \SetMathAlphabet\mathbfss{bold}{OT1}{cmss}{bx}{n}
  \ifAMStwofonts
    \ifCUPmtlplainloaded \else
      \DeclareSymbolFont{UPM}{U}{eur}{m}{n}
      \SetSymbolFont{UPM}{bold}{U}{eur}{b}{n}
      \DeclareSymbolFont{AMSa}{U}{msa}{m}{n}
      \DeclareMathSymbol{\upi}{0}{UPM}{"19}
      \DeclareMathSymbol{\umu}{0}{UPM}{"16}
      \DeclareMathSymbol{\upartial}{0}{UPM}{"40}
      \DeclareMathSymbol{\leqslant}{3}{AMSa}{"36}
      \DeclareMathSymbol{\geqslant}{3}{AMSa}{"3E}

      \let\leq=\leqslant 

    \fi
  \fi
\fi % End of NFSS release 2

\ifCUPmtlplainloaded \else
  \ifAMStwofonts \else % If no AMS fonts
    \def\upi{\pi}
    \def\umu{\mu}
    \def\upartial{\partial}
  \fi
\fi